# Adaptive Sparse Channel Estimation for Time-Variant MISO Communication Systems

Guan Gui, Wei Peng, Abolfazl Mehbodniya, and Fumiyuki Adachi
Department of Communication Engineering
Graduate School of Engineering,
Tohoku University
Sendai, Japan
{gui,peng,mehbod}@mobile.ecei.tohoku.ac.jp, adachi@ecei.tohoku.ac.jp

*Abstract*—Channel estimation problem is one of the key technical issues in time-variant multiple-input single-output (MSIO) communication systems. To estimate the MISO channel, least mean square (LMS) algorithm is applied to adaptive channel estimation (ACE). Since the MISO channel is often described by sparse channel model, such sparsity can be exploited and then estimation performance can be improved by adaptive sparse channel estimation (ASCE) methods using sparse LMS algorithms. However, conventional ASCE methods have two main drawbacks: 1) sensitive to random scale of training signal and 2) unstable in low signal-to-noise ratio (SNR) regime. To overcome these two harmful factors, in this paper, we propose a novel ASCE method using normalized LMS (NLMS) algorithm (ASCE-NLMS). In addition, we also proposed an improved ASCE method using normalized least mean fourth (NLMF) algorithm (ASCE-NLMF). Two proposed methods can exploit the channel sparsity effectively. Also, stability of the proposed methods is confirmed by mathematical derivation. Computer simulation results show that the proposed sparse channel estimation methods can achieve better estimation performance than conventional methods.

## I. INTRODUCTION

Signal transmission over multiple-input multiple-output (MIMO) channel is becoming one of the mainstream techniques in the next generation communication systems. The major motivation is due to the fact that MIMO technology is a way of using multiple antennas to simultaneously transmit multiple streams of data in wireless communication systems. MIMO in cellular systems brings improvements on four fronts: improved data rate, improved reliability, improved energy efficiency, and reduced interference. However, a coherent receiver requires accurate channel state information (CSI) due to the fact that wireless signal propagates over frequency-selective fading channel. In these systems, the basic channel estimation problem is to estimate the MISO channel at each antenna at receiving side. One of the typical examples is employing very large number of antenna (so-called "massive MIMO") at base station and only one antenna at mobile terminal (as shown in Fig. 1) to make high data communication possible with very low transmit power in a frequency-selective fading channel [1]. Besides, in a high mobility environment, the MISO channel is subjected to time-variant fading (i.e., double-selective fading). The accurate estimation of channel impulse response (CIR) is a crucial and challenging issue in coherent modulation and its accuracy has a significant impact on the overall performance of communication system.

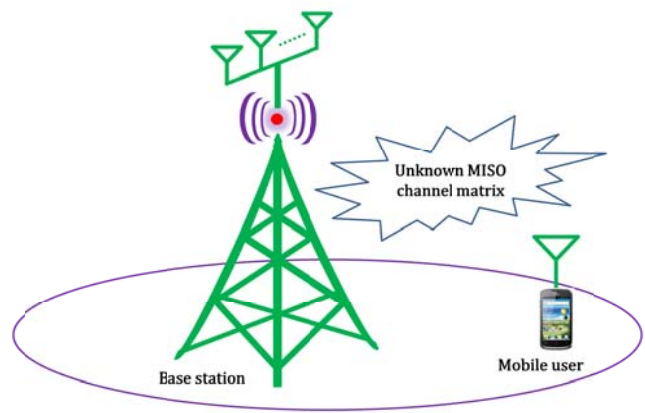

Fig.1. An example of time-variant MISO system.

In last decades, many channel estimation methods were proposed for MIMO-OFDM systems [2–10]. All these methods are categorized into two categories. The first category consists the linear channel estimation methods, e.g., least squares (LS) algorithm, which are based on the assumption of dense CIRs. The performance of linear methods depends only on the size of MIMO channel. Note that narrowband MIMO channel may be modeled as dense channel model because of its very short time delay spread; however, broadband MIMO channel is often modeled as sparse channel model [11–13]. A typical example of sparse channel is shown in Fig. 2. It is well known that linear channel estimation methods are relatively simple to implement due to their low computation complexity [4–9]. But, their main drawback is the failure to exploit the inherent channel sparsity. The second category consists the sparse channel estimation methods which use compressive sensing (CS) [14], [15]. Optimal sparse channel estimation often requires that its training signal satisfies restrictive isometry property (RIP) [16] in high probability. However, designing the RIP-satisfied training signal is a non-polynomial (NP) hard problem [17]. There exist some stable methods but with the cost of extra high computational burden, especially in time-variant MIMO-OFDM systems. For example, sparse channel estimation method using Dantzig selector was proposed for double-selective fading MIMO systems [9]. This method needs to be solved by linear programming, as a result incurs high computational complexity. To reduce complexity, sparse channel estimation methods using greedy iterative algorithms were also proposed in [8], [10]. However, their complexity depends on the number of nonzero taps of MIMO channel.

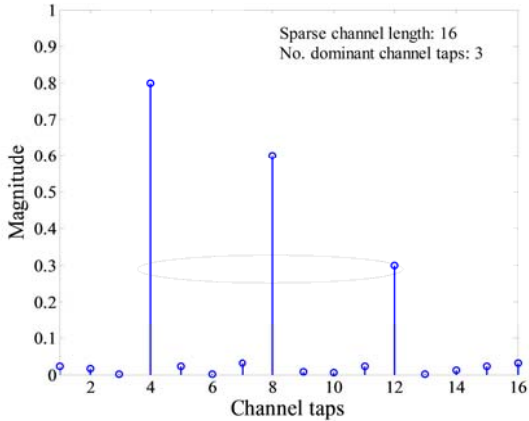

Fig. 2. A typical example of sparse multipath channel.

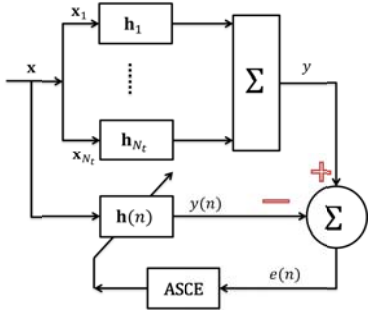

Fig. 3. ASCE for MISO-OFDM communication systems.

Unfortunately, the aforementioned methods cannot estimate the channel adaptively. To estimate time-variant channel, adaptive sparse channel estimation (ASCE) methods using sparse least mean square algorithms (ASEC-LMS) were proposed in [21]. However, conventional ASCE-LMS methods have two main drawbacks: 1) sensitive to random scale of training signal and 2) unstable in low signal-to-noise ratio (SNR) regime. To overcome the two harmful factors, in this paper, we propose a novel ASCE method using normalized LMS (NLMS) algorithm (ASCE-NLMS) for estimating MISO channel. Moreover, we also propose another improved ASCE method using sparse normalized least mean fourth (NLMF) algorithm (ASCE-NLMF) [18]. NLMF outperforms the well-known normalized least mean square (NLMS) algorithm [19] in achieving a better balance between complexity and estimation performances. In our previous research in [20], stable sparse NLMF algorithm was also proposed to achieve a better estimation than sparse NLMS algorithm [21]. Computer simulation results confirm the effectiveness of our proposed methods.

The remainder of this paper is organized as follows. A MISO-OFDM system model is described and problem formulation is given in Section II. In section III, sparse NLMS and sparse NLMF algorithms are introduced and ASCE in time-variant MISO-OFDM systems is highlighted. Computer simulation results are given in Section IV in order to evaluate and compare the performances of the proposed ASCE methods. Finally, we conclude the paper in Section V.

## II. SYSTEM MODEL AND PROBLEM FORMULATION

A time-variant MISO communication system using OFDM modulation, as shown in Fig. 1, is considered. Frequency-domain signal vector $\bar{\mathbf{x}}_{n_t}(t) = [\bar{x}_{n_t}(t,0), \ldots, \bar{x}_{n_t}(t, K-1)]^T$, $n_t = 1,2,\ldots,N_t$ is fed to inverse discrete Fourier transform (IDFT) at the $n_t$-th antenna, where $K$ is the number of subcarriers. Assume that the transmit power is $\{\|\bar{\mathbf{x}}_{n_t}(t)\|\} = KE_0$. The resultant vector $\mathbf{x}_{n_t}(t) \triangleq \mathbf{F}^H \bar{\mathbf{x}}_{n_t}(t)$ is padded with cyclic prefix (CP) of length $L_{CP} \geq (K-1)$ to avoid inter-block interference (IBI), where $\mathbf{F}$ is a $K \times K$ DFT matrix with entries $[\mathbf{F}]_{kq} = 1/K\, e^{-j2\pi kq/K}$, $k,q = 0,1,\ldots,K-1$. After CP removal, the received signal vector at the $n_t$-th antenna for time $t$ is written as $y$. Then, the received signal $y$ and input signal vector $\mathbf{x}$ are related by

$$y = \sum_{n_t=1}^{N_t} \mathbf{h}_{n_t}^T \mathbf{x}_{n_t} + z = \mathbf{h}^T \mathbf{x} + z, \quad (1)$$

where $\mathbf{x} = [\mathbf{x}_1^T, \mathbf{x}_2^T, \ldots, \mathbf{x}_{N_t}^T]^T$ combines all of the input signal vectors; additive noise variable $z$ satisfies $\mathrm{CN}(0, \sigma_n^2)$ and the MISO channel vector $\mathbf{h}$ can be written as

$$\mathbf{h} = [\mathbf{h}_1^T \quad \mathbf{h}_2^T \quad \cdots \quad \mathbf{h}_{N_t}^T]^T \in \mathbb{C}^{NN_t \times 1}, \quad (2)$$

where $\mathbf{h}_{n_t}$ ($n_t = 1,2,\ldots,N_t$) is assumed equal $N$-length sparse channel vector from receiver to $n_t$-th antenna. In addition, we also assume that each channel vector $\mathbf{h}_{n_t}$ is only supported by $T$ dominant channel taps. A typical example of 16-paths sparse multipath channel, which is supported by 3 dominant channel taps, is depicted in Fig. 2. According to the system model in Eq. (1), the corresponding channel estimation error $e(n)$ at time $t$ can be written as

$$e(n) = y - y(n) = y - \mathbf{h}^T(n)\mathbf{x}(n), \quad (3)$$

where $\mathbf{h}(n)$ denotes an adaptive MISO channel estimator of $\mathbf{h}$ and $y(n)$ is the output signal. A diagram of ASCE method for MISO-OFDM communication system was shown in Fig. 3. The goal of ASCE is to estimate MISO channel $\mathbf{h}$ using error signal $e(n)$ and input training signal $\mathbf{x}$. Traditional ASCE methods using sparse LMS algorithms were proposed to exploit channel sparsity. Their cost functions of ASCE methods can be concluded as

$$L_s(n) = \tfrac{1}{2} e^2(n) + \lambda_{slp} \|\mathbf{h}\|_p, \quad (4)$$

where $0 \leq p < 1$ and $\lambda_{slp} \geq 0$ denotes the sparse regulation parameter which trades off the mean square error and sparsity of **h**. Without lose generality, corresponding update equation of ASCE methods can be written as

$$\mathbf{h}(n+1) = \mathbf{h}(n) - \mu_s \frac{\partial L_s(n)}{\partial \mathbf{h}(n)}$$
$$= \mathbf{h}(n) + \mu_s e(n)\mathbf{x}(n) - \rho_{slp} \frac{\|\mathbf{h}(n)\|_p^{1-p} \text{sgn}(\mathbf{h}(n))}{\sigma + |\mathbf{h}(n)|^{1-p}}, \quad (5)$$

where $\rho_{slp} = \mu_s \lambda_{slp}$ and $\mu_s \in (0, \gamma_{\max}^{-1})$ is the step size of LMS gradient descend and $\gamma_{\max}$ is the maximum eigenvalue of the covariance matrix $\mathbf{R} = E\{\mathbf{x}(t)\mathbf{x}^T(t)\}$.

### III. PROPOSED ASCE METHODS

#### A. ASCE-NLMS

We consider a $L_p$-norm sparse penalty on cost function of NLMS to produce a sparse channel estimator, since this penalty term forces the channel taps values of **h** to approach zero. It is termed as LP-NLMS which was proposed for single-antenna systems in [21]. According to (4) and (5), update equation of LP-NLMS based ASCE method is given by

$$\mathbf{h}(n+1) = \mathbf{h}(n) + \mu_s \frac{e(n)\mathbf{x}(n)}{\|\mathbf{x}(n)\|_2^2} - \rho_{slp} \frac{\|\mathbf{h}(n)\|_p^{1-p} \text{sgn}(\mathbf{h}(n))}{\sigma + |\mathbf{h}(n)|^{1-p}}. \quad (6)$$

where $\|\cdot\|_2$ is the Euclidean norm operator and $\|\mathbf{x}\|_2^2 = \sum_{i=1}^N |x_i|^2$. If $p = 0$ in Eq. (4), then it is termed as $L_0$-norm NLMS (L0-NLMS) [21] and the cost function is given by

$$L_{sl0}(n) = \frac{1}{2}e^2(n) + \lambda_{sl0}\|\mathbf{h}(n)\|_0, \quad (7)$$

where $\|\mathbf{h}\|_0$ is the $L_0$-norm operator that counts the number of nonzero taps in **h** and $\lambda_{sl0}$ is a regularization parameter to balance the estimation error and sparse penalty. Since solving the $L_0$-norm minimization is a NP-hard problem [17], we replace it with an approximate continuous function

$$\|\mathbf{h}\|_0 \approx \sum_{l=0}^{N_tN-1}(1 - e^{-\beta|h_l|}). \quad (8)$$

According to the approximate function, L0-LMS cost function can be revised as

$$L_{sl0}(n) = \frac{1}{2}e^2(n) + \lambda_{sl0}\sum_{l=0}^{N_tN-1}(1 - e^{-\beta|h_l|}). \quad (9)$$

Then, the update equation of L0-NLMS based ASCE can be derived as

$$\mathbf{h}(n+1) = \mathbf{h}(n) + \mu_s e(n)\mathbf{x}(n)$$
$$-\rho_{sl0}\beta \text{sgn}(\mathbf{h}(n))e^{-\beta|\mathbf{h}(n)|}, \quad (10)$$

where $\rho_{sl0} = \mu_s \lambda_{sl0}$. It is worth mentioning that the exponential function in (10) has high computational complexity. To reduce the computational complexity, the first order Taylor series expansion of exponential functions is considered as [22]

$$e^{-\beta|h|} \approx \begin{cases} 1 - \beta|h|, & \text{when } |h| \leq 1/\beta \\ 0, & \text{others.} \end{cases} \quad (11)$$

where $h$ is any element of channel vector **h**. Finally, the update equation of L0-NLMS based adaptive sparse channel estimation can be derived as

$$\mathbf{h}(n+1) = \mathbf{h}(n) + \mu_s \frac{e(n)\mathbf{x}(n)}{\|\mathbf{x}(n)\|_2^2} - \rho_{l0}J(\mathbf{h}(n)), \quad (14)$$

where $J(\mathbf{h})$ is defined by

$$J(\mathbf{h}) = \begin{cases} 2\beta^2 h - 2\beta \text{sgn}(h), & \text{when } |h| \leq 1/\beta \\ 0, & \text{others.} \end{cases} \quad (15)$$

#### B. ASCE-NLMF

Unlike the proposed method in Section A, here we propose a kind of improved ASCE method using sparse NLMF algorithm for MISO channel. First, a cost function $L_{nlmf}(n)$ of standard LMF can be constructed as

$$L_f(n) = \frac{1}{4}e^4(n). \quad (16)$$

The update equation of ASCE using LMF algorithm is given by

$$\mathbf{h}(n+1) = \mathbf{h}(n) - \mu_f \frac{\partial L_{nlmf}(n)}{\partial \mathbf{h}(n)}$$
$$= \mathbf{h}(n) + \mu_f e^3(n)\mathbf{x}(n), \quad (17)$$

where $\mu_f \in (0,2)$ is a gradient descend step-size which controls convergence speed and steady-state performance; However, LMF algorithm only works stable in low SNR regimes [23]. Based on our previous research in [21], ASCE-NLMF algorithm is stable for different SNR regimes. The update equation of ASCE-NLMF is derived as

$$\mathbf{h}(n+1) = \mathbf{h}(n) + \mu_f \frac{e^3(n)\mathbf{x}(n)}{\|\mathbf{x}(n)\|_2^2(\|\mathbf{x}(n)\|_2^2 + e^2(n))}$$
$$= \mathbf{h}(n) + \mu_f(n)\frac{e(n)\mathbf{x}(n)}{\|\mathbf{x}(n)\|_2^2}, \quad (18)$$

where $\mu_f(n) = \mu_f e^2(n)/(\|\mathbf{x}(n)\|_2^2 + e^2(n))$. Here, we observe that when $e^2(n) \gg \|\mathbf{x}(n)\|_2^2$, then $\mu_f(n) \to \mu_f$; when $e^2(n) \approx \|\mathbf{x}(n)\|_2^2$, then $\mu_f(n) \to \mu_f/2$; when $e^2(n) \ll \|\mathbf{x}(n)\|_2^2$, then $\mu_f(n) \to 0$. Hence, NLMF algorithm in Eq. (18) is stable which is equivalent to NLMS algorithm in Eq. (6). According to the previous research in [21], if the standard NLMF algorithm is stable, then its corresponding ASCE method using sparse NLMF algorithm is also stable.

For MISO channel vector **h**, the cost function of ASCE using LP-NLMF algorithm is given by

$$L_{flp}(n) = \frac{1}{4}e^4(n) + \lambda_{flp}\|\mathbf{h}\|_p, \quad (19)$$

where $\lambda_{flp}$ is a regularization parameter which trades off the fourth-order mismatching estimation error and $L_p$-norm sparse penalty of **h**. The update equation of ASCE method using LP-NLMS can be derived as

$$\mathbf{h}(n+1) = \mathbf{h}(n) + \mu_f(n)\frac{e(n)\mathbf{x}(n)}{\|\mathbf{x}(n)\|_2^2} - \rho_{flp}\frac{\|\mathbf{h}(n)\|_p^{1-p}\text{sgn}(\mathbf{h}(n))}{\sigma + |\mathbf{h}(n)|^{1-p}}, \quad (20)$$

where $\rho_{flp} = \mu_f \lambda_{flp}$ depends on gradient descend step-size $\mu_f$ and regularization parameter $\lambda_{flp}$. Similarly, cost function of ASCE method using L0-LMF algorithm is written as

$$L_{fl0}(n) = \frac{1}{4}e^4(n) + \lambda_{fl0}\|\mathbf{h}(n)\|_0, \quad (21)$$

where $\lambda_{fl0} > 0$ is a regularization parameter which trades off the fourth-order mismatching estimation error and sparseness

of MISO channel. The corresponding updating equation algorithm is given by

$$\mathbf{h}(n+1) = \mathbf{h}(n) + \mu_f(n)\frac{e(n)\mathbf{x}(n)}{\|\mathbf{x}(n)\|_2^2} - \beta_2 J(\mathbf{h}(n)), \quad (22)$$

where $\beta_2 = \mu_f \lambda_{fl0}$ and $J(\mathbf{h})$ is an approximate sparse $L_0$-norm function, defined in Eq. (15).

## IV. NUMERICAL SIMULATIONS

In this section, we evaluate our proposed ASCE estimators using 1000 independent Monte-Carlo runs for averaging. The length of channel vector $\mathbf{h}_{n_t}$ between each transmitter and receiver antenna is set to $N = 16$ and the number of dominant taps is set to $T = 1$ and $3$, respectively. Values of dominant channel taps follow Gaussian distribution and their positions are randomly allocated within the length of $\mathbf{h}_{n_t}$ which is subjected to $E\{\|\mathbf{h}_{n_t}\|_2^2 = 1\}$. The received signal-to-noise ratio (SNR) is defined by $20\log(E_0/\sigma_n^2)$, where $E_0 = 1$ is the transmit power at each antenna. Here, we set the SNR as 3dB, 6dB and 9dB in computer simulation. All of the step sizes and regularization parameters are listed in Table I. The estimation performance is evaluated by average mean square error (MSE) which is defined as

$$\text{Avergae MSE}\{\mathbf{h}(n)\} = E\{\|\mathbf{h} - \mathbf{h}(n)\|_2^2\}, \quad (23)$$

where $E\{\cdot\}$ denotes the expectation operator, $\mathbf{h}$ and $\mathbf{h}(n)$ are the actual MISO channel vector and its $n$-th adaptive channel estimator, respectively.

TABLE I. SIMULATION PARAMETERS.

| Parameters | Values |
|---|---|
| Gradient descend step-size: $\mu_s$ | 0.5 |
| Gradient descend step-size: $\mu_f$ | 1.5 |
| Regularization parameter: $\lambda_{slp}$ | $(2e-4)\sigma_n^2 \log(N/T)$ |
| Regularization parameter: $\lambda_{flp}$ | $(2e-6)\sigma_n^2 \log(N/T)$ |
| Regularization parameter: $\lambda_{sl0}$ | $(2e-3)\sigma_n^2 \log(N/T)$ |
| Regularization parameter: $\lambda_{fl0}$ | $(2e-5)\sigma_n^2 \log(N/T)$ |

In the first example, the proposed methods are evaluated in Fig. 4 ($T = 1$) and Fig. 5 ($T = 3$) at SNR = 3dB. To balance the estimation performance and computational complexity, the step-size of sparse NLMS algorithms and sparse NLMF algorithms are set as $\mu_s = 0.5$ and $\mu_f = 1.5$, respectively. Note that the step-size $\mu_s = 0.5$ was also recommended by the paper [21]. As the two figures show, ASCE-NLMS method achieves better estimation performance than ACE-NLMS. Similarly, ASCE-NLMF method also achieves better estimation performance than ACE-NLMF method. We can also observe that ASCE-NLMF method outperforms the ASCE-NLMS method significantly but with the cost of higher computational complexity (iterative times). Relatively, the computational complexity of ASCE-NLMS is very low [21]. As a result, selecting a reasonable ASCE method depends on the requirements of the system to be designed.

In the second experiment, the proposed methods are evaluated at SNR regimes 6dB and 9dB as shown in Figs. 6-7.

Again, we can notice improvement over conventional methods. Please note that computational complexity of ASCE-NLMF method increases with SNR. Our future work is finding a solution for reducing the complexity in ASCE-NLMF.

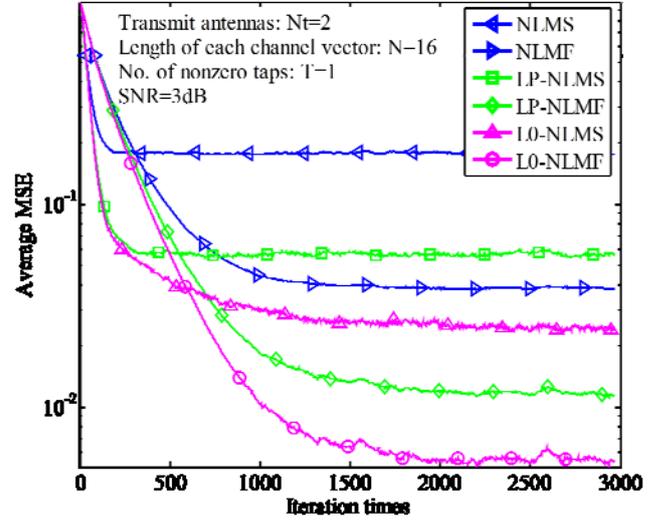

Fig. 4. Performance comparison at SNR = 3dB.

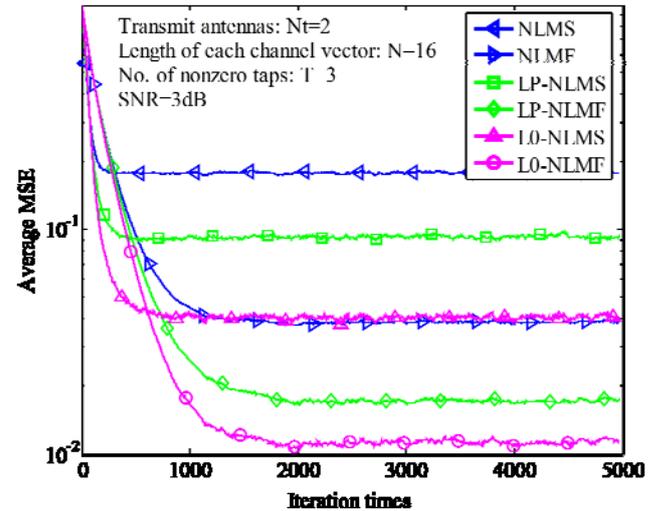

Fig. 5. Performance comparison at SNR = 3dB.

## V. CONCLSION

In this paper, we proposed a ASCE method using sparse NLMS and sparse NLMF algorithms for time-variant MISO-OFDM systems. First of all, system model was formulated to ensure each MISO channel vector can be estimated. Secondly, cost functions of the two proposed methods were constructed using sparse penalties, i.e., $L_p$-norm and $L_0$-norm. Later, MISO channel vector was estimated using ASCE method. Simulation results indicate that the proposed ASCE-NLMS method achieves a better performance than the standard ACE-NLMS method without much increase in computational complexity. The simulation results also demonstrate that the proposed ASCE-NLMF method is even better than ASCE-NLMS method but has a higher amount of computation complexity.

ACKNOWLEDGMENT

This work was supported in part by the Japan Society for the Promotion of Science (JSPS) postdoctoral fellowship.

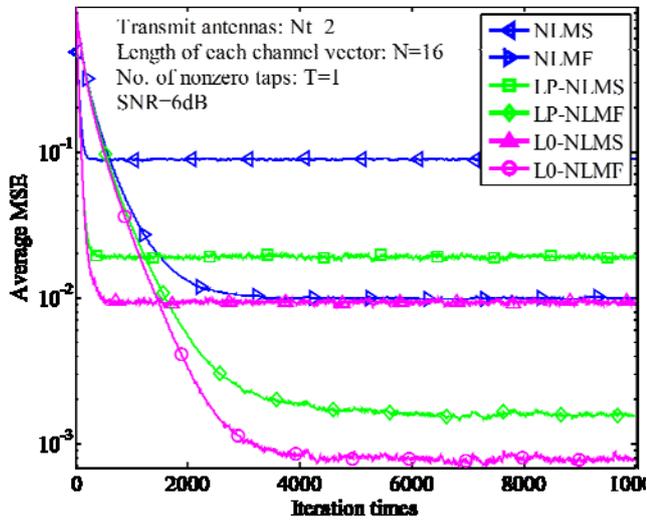

Fig. 6. Performance comparison at SNR = 6dB.

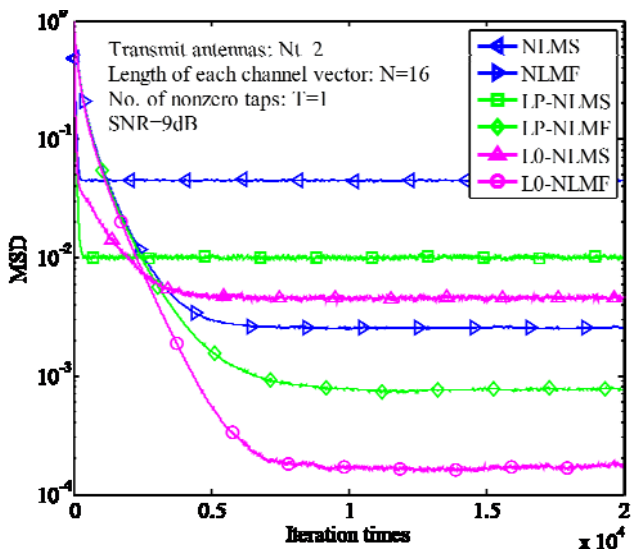

Fig. 7. Performance comparison at SNR = 9dB.